\documentclass[aps,twocolumn,nofootinbib,superscriptaddress,amsfonts,floatfix]{revtex4-1} 
\usepackage{graphicx,amsmath,amssymb,amstext}
\usepackage{amssymb,amsbsy,amsfonts,amsthm,color}

\usepackage{epsfig}
\usepackage{graphicx}
\usepackage{subfigure}
\usepackage{hyperref}
\usepackage{cancel}
\usepackage{balance}
\usepackage{float}
\usepackage{ulem}

\graphicspath{{Figures/}}

\usepackage{color}
\usepackage[dvipsnames]{xcolor}

\usepackage{comment}

\newcommand{\mpl}{m_\mathrm{Pl}}
\newcommand{\mbh}{M_\mathrm{BH}}

\newcommand{\grchombo}{\mathtt{GRChombo}}
\newcommand{\be}{\begin{equation}}
\newcommand{\ee}{\end{equation}}

\newcommand{\eqn}[1]{Eqn. (\ref{#1})}

\hypersetup{
  colorlinks   = true, 
  urlcolor     = blue, 
  linkcolor    = blue, 
  citecolor   = teal 
}

\newcommand{\Mxi}{M^0_\xi}
\newcommand{\axi}{a^0_\xi}

\begin{document}
{\hfill KCL-PH-TH/2023-35}
\title{Spinning primordial black holes formed during a matter-dominated era}

\author{Eloy de Jong}
\email{eloy.dejong@kcl.ac.uk}
\affiliation{Theoretical Particle Physics and Cosmology Group, Physics Department, Kings College London, Strand, London WC2R 2LS, United Kingdom}

\author{Josu C. Aurrekoetxea}
\email{josu.aurrekoetxea@physics.ox.ac.uk}
\affiliation{Astrophysics, University of Oxford, Denys Wilkinson Building, Keble Road, Oxford OX1 3RH, United Kingdom}

\author{Eugene A. Lim}
\email{eugene.a.lim@gmail.com}
\affiliation{Theoretical Particle Physics and Cosmology Group, Physics Department, Kings College London, Strand, London WC2R 2LS, United Kingdom}

\author{Tiago Fran\c{c}a}
\email{t.e.franca@qmul.ac.uk}
\affiliation{School of Mathematical Sciences, Queen Mary University of London, Mile End Road, London E1 4NS, United Kingdom}

\begin{abstract}    
    We study the formation of spinning primordial black holes during an early matter-dominated era. Using non-linear 3+1D general relativistic simulations, we compute the efficiency of mass and angular momentum transfer in the process -- which we find to be $\mathcal{O}(10\%)$ and $\mathcal{O}(5\%)$, respectively. We show that subsequent evolution is important due to the seed PBH accreting non-rotating matter from the background, which decreases the dimensionless spin.  Unless the matter era is short, we argue that the final dimensionless spin will be negligible. 
\end{abstract}
\maketitle

\section*{Introduction} \label{Ssect:intro}

The realization that cold dark matter may be partly made up out of primordial black holes (PBHs) \cite{Carr:1974, Hawking:1971,Zeldovich:1967,Carr:2016ks, Carr:2017rt, Carr:2020xqk} and the opportunities to test and constrain this hypothesis \cite{Page:1976wx,Carr:1976zz,Wright:1995bi,Lehoucq:2009ge,Kiraly1981,MacGibbon:1991vc,Cline:1996zg, Carr:1984cr,Bean:2002kx, Hawkins:1993,Carr:2019kxo,Hawkins:2020, LIGOScientific:2020ufj,Franciolini:2021tla, DeLuca:2020agl,Vaskonen:2020lbd,Kohri:2020qqd,Domenech:2020ers, Arzoumanian_2020,Atal:2022zux,Carr:2023tpt} has renewed interest in PBHs. The mechanism considered standard to form these relics is via the collapse during the radiation era of superhorizon perturbations originating from the growth of quantum fluctuations during inflation \cite{Carr:1975,Nadezhin:1978,Bicknell:1979, Choptuik:1993,Evans:1994,Niemeyer_1998:nj,Green_1999:gl,Musco:2012au,Yoo:2020lmg,Carr:1993cl,Carr_1994:cgl,Hodges:1990hb,Ivanov:1994inn,Garcia:1996glw,Randall:1996rsg,Taruya_1999,Bassett_2001,Clesse_2015,Inomata_2017,Garc_a_Bellido_2017,Ezquiaga:2017fvi,Geller:2022nkr,Qin:2023lgo}, but could also be the result of other early universe dynamical processes or epochs \cite{Crawford1982,Hawking:1982,Kodama:1982,Leach:2000ea,Moss:1994,Kitajima:2020kig,Khlopov:1998nm,Konoplich:1999qq,Khlopov:1999ys,Khlopov:2000js,Kawana:2021tde,Jung:2021mku,Dokuchaev:2004kr,Rubin:2000dq,Rubin:2001yw,Garriga:2015fdk,Deng:2016vzb,Liu:2019lul, Hogan:1984zb,Hawking:1987bn,Polnarev:1991,Garriga:1993gj,Caldwell:1995fu,
MacGibbon:1997pu,Wichoski:1998ev,Hansen:1999su,Nagasawa2005,Carr:2009jm,
Bramberger:2015kua,Helfer:2019,Bertone:2019irm,James-Turner:2019ssu,Aurrekoetxea:2020tuw}. 

There is next to no observational data to constrain the thermal history of our universe before big bang nucleosynthesis \cite{Carroll:2001bv,Hooper:2023brf}. Models of early matter-dominated expansion epochs include overproduction of non-relativistic particles \cite{Khlopov:1980mg,Polnarev:1982a}, the presence of moduli fields ubiquitous in string theory models \cite{Green:1997,Kane:2015jia}, or at the end of inflation during a process known as reheating \cite{Kofman:1994rk, Kofman:1997yn,Albrecht:1982mp,Amin:2014eta,Aurrekoetxea:2023jwd,Carr:2018}, when the energy stored in the inflaton is transferred into standard model particles.

PBHs' properties and abundances, and therefore corresponding detection prospects, depend on the details of the era in which they form. In this work, we investigate the expected angular momentum of PBHs. Angular momentum is relevant to possible PBH signatures, e.g. the amplitude of the stochastic gravitational wave background from spinning PBHs could increase by 50\% \cite{Kuhnel:2019zbc}, and PBHs with spin may avoid some of the abundance bounds related to evaporation due to their lower Hawking temperatures \cite{Arbey:2019jmj}. It is suggested that PBHs that form during a radiation-dominated era generally have small spins \cite{DeLuca:2019buf, Mirbabayi:2019uph, Harada:2020pzb,Chongchitnan:2021ehn}, reflected by the fact that the PBH formation threshold is increased proportional to the square of the angular momentum \cite{He:2019cdb}, although they could develop non-negligible spins through Hawking radiation \cite{Calza:2021czr,Calza:2023rjt}. On the other hand, PBH production is more efficient in a matter-dominated era in the absence of pressure support, even though non-spherical effects that may resist gravitational collapse become important \cite{Harada:2016mhb}. PBH formation in the context of an early matter-dominated epoch has also been studied in e.g. \cite{Green:1997jkl,Cotner:2016cvr,Hidalgo:2017dfp,Georg:2016yxa,Georg:2017mqk,Carr:2017edp,Kokubu:2018fxy,DeLuca:2021pls,Padilla:2021zgm,Harada:2022xjp,Hidalgo:2022yed}. Of particular relevance to this work is the argument by the authors of \cite{Harada:2017fjm} that most PBHs formed in a matter-dominated setting are near-extremal, although this result does not take into account PBH mass accretion.\\

In this work, we report on fully non-linear 3+1D numerical relativity simulations that aim to shed more light on the role of angular momentum in PBH formation during a matter-dominated era. We simulate the collapse of superhorizon non-linear perturbations sourced by a massless scalar field, on a matter-dominated expanding background driven by an oscillating massive scalar field.

\begin{figure*}[t]
    \href{https://youtu.be/CC4xBLol4aE}{
    \includegraphics[width=\linewidth]{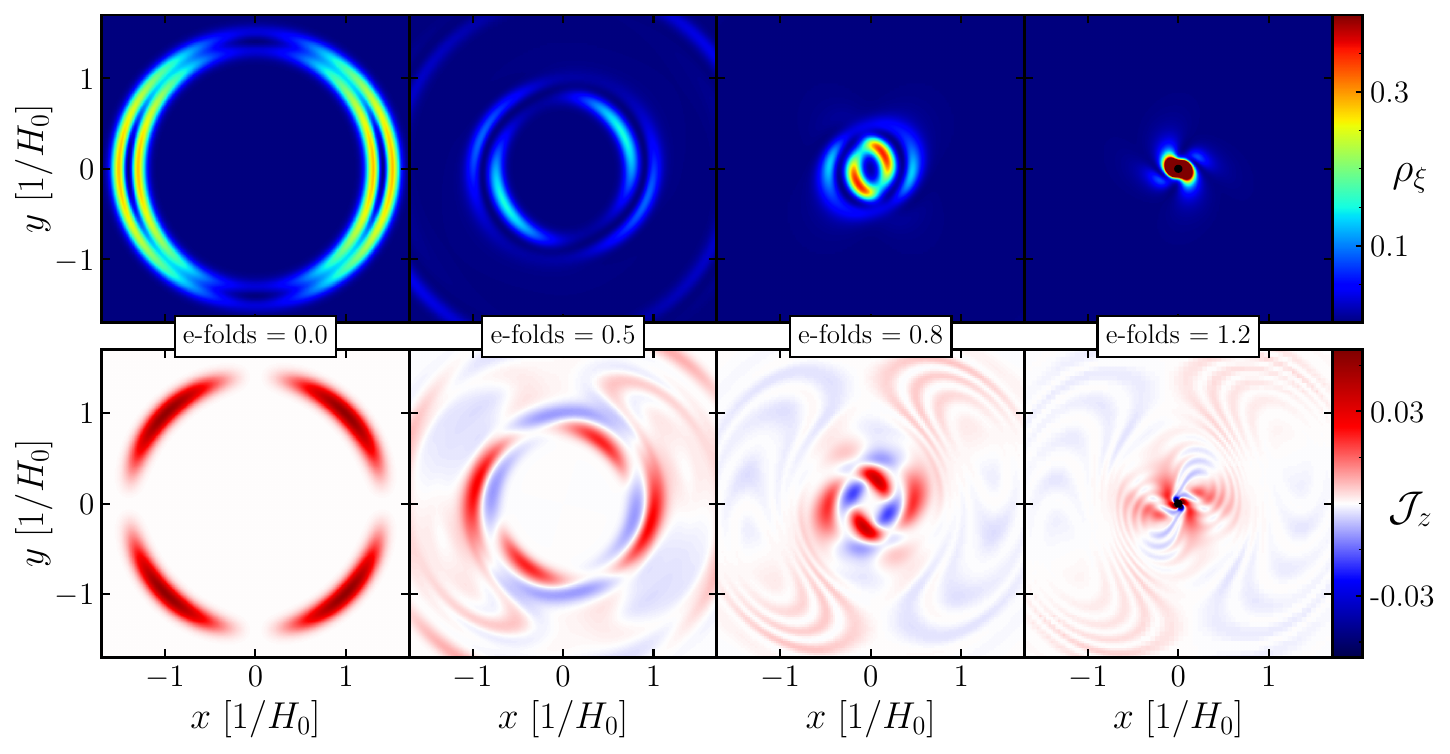}}
    \caption{\textbf{Evolution panel} showing the evolution of the energy density of the field $\xi$ (total angular momentum density around the $z$ axis) in the top (bottom) row, in units of $[\rho_{\xi}]=H_0^2 \mpl^2$ ($[\mathcal{J}_z]=H_0 \mpl^2$), for a perturbation with $R_0 = 1.4 H_0^{-1}$. The leftmost column shows the initial conditions, in which $\rho_\xi$ shows two peaks corresponding to the gradients of the gaussian in \eqn{Seq::xi_r_init}, whilst the angular momentum density takes strictly positive values. In the top row, we see the perturbation collapse and acquire an asymmetric configuration before black hole formation in the third column, signalling the presence of angular momentum in the collapsing region. In the bottom panels, we see $\mathcal{J}_z$ take on both positive and negative values over the course of the evolution, on which we comment in section \ref{Ssect::results_pbhspin}. In the rightmost column the PBH has formed, and the small area already hidden behind the apparent horizon is shown in black. A video of this data can be found \href{https://youtu.be/CC4xBLol4aE}{here} \cite{Movie3}.
    } 
    \label{Sfig:evol_panel}
\end{figure*}

We find that the formation process is quite efficient, i.e. at horizon formation the PBH contains $\mathcal{O}(10\%)$ of the collapsing overdensity's mass and angular momentum. However, the PBH dimensionless spin goes down as it accretes non-rotating background matter, resulting in negligible final spins if the matter-dominated era lasts several e-folds. We illustrate typical collapse behaviour in Fig. \ref{Sfig:evol_panel}. Note that since a radiation-dominated era must  intervene before the  onset of the present matter-dominated era, the background massive scalar must reheat. Any PBHs that were formed during this era will remain as a matter component which presumably can be the dark matter component today.\\

The paper is organised as follows. In section \ref{Ssect:initialconditions}, we discuss the numerical setup of our simulations, initial conditions and diagnostics. In particular, we provide details on finding black hole apparent horizons in expanding spacetimes. In section \ref{Ssect:results} we report on our main results, after which we conclude in section \ref{Ssect:conclusions}.

\vspace*{\fill} 

\section{Setup} \label{Ssect:initialconditions}

We consider a massive scalar field $\phi$ and a massless scalar field $\xi$, both minimally coupled to gravity\footnote{We use the mostly plus $-+++$ signature  and Planck units $\hbar=c=1$, such that $G=\mpl^{-2}$, where $\mpl$ is the non-reduced Planck mass.}, i.e.
\begin{equation}
S = \int ~d^4 x \sqrt{-g}\left[\frac{\mpl^2}{16\pi}R - \mathcal{L}_\phi - \mathcal{L}_\xi\right]~.
\end{equation}
We also assume there is no direct coupling between them, i.e.
\begin{subequations}
\begin{align}
\mathcal{L}_\phi &=\frac{1}{2}\nabla^{\mu}\phi\nabla_\mu\phi + \frac{1}{2}m^2 \phi^2~, \\
\mathcal{L}_\xi &= \frac{1}{2}\nabla^{\mu}\xi\nabla_\mu\xi ~.
\end{align}
\end{subequations}

The scalar field $\phi$ with mass $m$ plays the role of the ambient matter driving the background expansion. In the limit in which $\phi$ is homogeneous and dominates, the spacetime is well-described by the Friedman-Lema\^{i}tre-Robertson-Walker (FLRW) line element
\begin{equation}
    ds^2 = -dt^2 + a(t)^2(dr^2+r^2d\Omega^2_2)\,.
\end{equation}
If $\phi$ oscillates coherently with a period considerably smaller than a Hubble time, $2\pi/m \ll 1/H$, its pressure averages to zero and the scale factor $a(t)$ grows in a matter-like fashion, i.e. $a \propto t^{2/3}$ and $\rho\propto a^{-3}$.
The massless field $\xi$, on the other hand, plays the role of the collapsing superhorizon perturbation, and will thus be the seed triggering black hole formation. 

Going beyond the FLRW limit to study black hole formation requires solving the Einstein field equations \cite{Einstein:1916} for a more general line element, which we decompose in the usual ADM form \cite{Arnowitt_2008}
\begin{equation}
    ds^2 = -\alpha^2dt^2 + \gamma_{ij}(dx^i + \beta^i dt)(dx^j + \beta^j dt),
\end{equation}
where $\gamma_{ij}$ is the three-dimensional spatial metric. The lapse and shift gauge functions $\alpha$ and $\beta^i$ determine the choice of spatial hypersurface and their coordinates, which in numerical relativity are dynamically determined. At each hypersurface, the rate of change of $\gamma_{ij}$ is given by the extrinsinc curvature $K_{ij} \equiv -(1/2){\cal{L} }_{{\bf n}}\gamma_{ij}$, where ${\bf n}$ is the vector normal to the hypersurface. $K_{ij}$ can be decomposed further into its trace $K$ and tracefree components $A_{ij}$, where $K_{ij} = A_{ij} + (1/3)K\gamma_{ij}$. 
The trace $K$ measures the local expansion, where in our sign convention negative (positive) $K$ indicates locally expanding (collapsing) space.
 
We evolve the Einstein field equations using the CCZ4 formulation \cite{Alic_2012} and the moving puncture gauge \cite{Bona:1994dr,Baker:2005vv,Campanelli:2005dd,vanMeter:2006vi}, using the numerical relativity code $\grchombo$ \cite{Clough:2015sqa,Radia:2021smk,Andrade2021}. We list explicit expressions for the matter variable evolution equations in appendix \ref{Sapp::evolution_equations}.

\subsection{Initial data}\label{Ssect::init_data}

We choose the initial gauge $\alpha=1$ and $\beta^i=0$.
We set the scalar field $\phi$ to a homogeneous value $\phi_0$ starting from rest, i.e. $\partial_t{\phi}=0$. The initial unperturbed Hubble parameter $H_0$ is then 
\begin{equation} \label{Seq::initial_hubble_param}
    H_0^2 = \frac{4\pi m^2}{3\mpl^2}\phi_0^2 \,.
\end{equation}
We break spherical symmetry and inject angular momentum into the system via the elevation and azimutal angle-dependence\footnote{Note the difference between $\varphi$, which denotes the azimuthal angle of a spherical coordinate system, and $\phi$, denoting the massive scalar field.} of the massless field $\xi$ and its conjugate momentum $\Pi_\xi=\left(\partial_t \xi -\beta^i\partial_i\xi\right)/\alpha$
\begin{subequations}
\begin{align}
    \label{Seq::xi_init_time} \xi(t,r,\theta,\alpha) &= R(r)\left[1 + \sin{(\theta)}\Phi(t,\varphi)\right],\\
    \label{Seq::pi_init_time} \Pi_\xi (t,r,\theta,\alpha) &=R(r)\sin{(\theta)}\partial_t\Phi(t,\varphi),
\end{align}
\end{subequations}
where
\begin{subequations}
    \begin{align}
        \label{Seq::xi_r_init} R(r) &= A \exp{\left[-\frac{(r-R_0)^2}{\lambda^2}\right]},\\
        \label{Seq::xi_phi_init} \Phi(t,\varphi) &= B \cos{(k\varphi-\omega t)}.
    \end{align}
\end{subequations}
The constants $A,\, R_0,\, \lambda,\, B,\, k,\, \omega$ represent the perturbation shell's initial radial amplitude, radius, width, spin amplitude, spin wavenumber and spin angular velocity, respectively -- Fig. 
\ref{Sfig:init_params} in appendix \ref{Sapp::initial_data} schematically clarifies the meaning of these constants.
The angular momentum density around any given axis through the origin is given by 
\begin{equation}
    \mathcal{J}^i = \epsilon^{ijk}x_j S_k
\end{equation}
where $S^i = -\gamma^{ij}n^a T_{aj}$ and $n^a = (1,-\beta^i)/\alpha$ is the normal vector to the hypersurface. Here $\epsilon^{ijk}$ is the antisymmetric Levi-Civita symbol and $i,j,k=1,2,3$ label the spatial Cartesian basis.
We are interested in the $z$-component of the angular momentum density\footnote{The $\theta$-dependence introduces additional angular momentum around the $x$- and $y$-axes, but we choose parameters that makes sure these are negligible.}, which in the initial data is only sourced by the massless field $\xi$ and in a non-orthonormal polar basis $(\partial_r, \partial_\varphi,\partial_z)$ is expressed as
\begin{equation}\label{Seqn::angmomdensity_z}
\begin{aligned}
    \mathcal{J}_{\xi,z}^0 &= xS_{\xi,y} - yS_{\xi,x}\\
    &= -\Pi_{\xi}\left(x\partial_y\xi - y\partial_x\xi\right)\\
    &= -\Pi_{\xi} \partial_\varphi\xi,
\end{aligned}
\end{equation}
where we take a derivative with respect to the azimuthal angle $\varphi$ directly in the last line and all quantities on the RHS are evaluated on the initial slice. This expression can be integrated over the volume to find the total initial angular momentum $J_{\xi,z}^0$. 
In our notation, the $\mathcal{J}$ in a calligraphic font will denote an angular momentum density, while $J$ will denote angular momentum, i.e. $J_i = \int ~dV \mathcal{J}_i$. In what follows, if a directional subscript is omitted, a $z$-subscript is implied. 
Finally, we note the important point that the angular momentum is \emph{not} concentrated on the equatorial plane of the shell, but is roughly Gaussian distributed around it with angular width $\Delta \theta \approx 0.5\pi$ (see Fig. \ref{Sfig:theta_plot}). The implications of this distribution is that, as we shall see, matter which is spinning but not along the equatorial plane will still spiral into the center. 

As is well known, all initial data in general relativity must obey a coupled system of Hamiltonian and momentum constraint equations, which we solve using the CTTK method \cite{Aurrekoetxea:2022mpw}. We choose an initial spatially flat metric $\gamma_{ij}=\delta_{ij}$ and solve for the trace $K$ and traceless parts $A_{ij}$ of the extrinsic curvature tensor. This choice is equivalent to choosing an initially homogeneous cosmological scale factor, where the rate of local expansion is determined by the matter distribution. Note that in the absence of inhomogeneities, the Hamiltonian constraint reduces to the usual form of the Friedmann equation $H^2 = 8\pi\rho/3\mpl^2$, with $K=-3H$.

\subsection{Diagnostics quantities}

The key three diagnostic quantities we will track are the PBH mass $M_\mathrm{BH}$, dimensionful angular momentum $J_\mathrm{BH}$ and dimensionless spin (dimensionless Kerr parameter)
\begin{equation}
    a_\mathrm{BH}\equiv \frac{J_\mathrm{BH}}{G\mbh^2}~.
\end{equation}
We obtain these values by measuring the properties of the black hole \emph{apparent} horizon (AH) --  the outermost marginally trapped surface on which the expansion of outgoing null geodesics vanishes. To find the AH we follow the procedure described in \cite{Thornburg:2003sf} -- the next paragraph is rather technical and the reader may skip it. 

If the spatial metric on a given three dimensional spatial hypersurface is $\gamma_{ij}$, a two dimensional surface $S$ with a spacelike unit outward-pointing normal vector $s^a$ induces a two dimensional metric $m_{ab}$ on $S$ equal to
\begin{equation}
    m_{ab} = \gamma_{ab} - s_a s_b.
\end{equation}
The expansion of outgoing null vectors $k_+^{~a} = n^a + s^a$ in this surface is defined as $\Theta_+ \equiv m_{ab}\nabla^a k_+^{~b}$, or equivalently
\begin{equation} \label{Seqn::expansion_outgoing}
    \Theta_+\equiv D_i s^i + K_{ij} s^i s^j - K,
\end{equation}
which vanishes on the AH.
To find this surface during the formation stage, we shoot rays in different directions from a PBH centre \emph{guess} point, which we take to be the location of the maximum energy density, and solve $r^\kappa\Theta_+(r) = 0$ for $r$ along these rays. Here $r$ is coordinate distance along a given ray and $\kappa\geq 0$ is an arbitrary AH expansion radius power that can be chosen to optimise AH searches depending on underlying spacetimes. We find that in expanding spacetimes, $\kappa = 2$ provides reliable performance. 

To measure the area, spin and mass of the AH, we use \cite{Caudill:2006hw, Ashtekar:2000hw}
\begin{subequations}
\begin{align}
    A_\textrm{BH} &= \oint_{S} d^2 V,\\
    J^{i}_\textrm{BH} &= \frac{1}{8\pi} \oint_S ~ d^2 V \left(\xi^i\right)^l s^j K_{jl},\\
    M_\textrm{BH} &= \sqrt{\frac{A_\textrm{BH}}{16\pi} + \frac{4\pi J_\textrm{BH}^2}{A_\textrm{BH}}},
\end{align}
\end{subequations}
where $d^2V$ is the natural area measure on $S$ constructed from $m_{ab}$ and $\xi^i$ are Killing vectors of $m_{ab}$ approximated well by the flat space rotational Killing vectors, $\left(\xi^i\right)^l=\epsilon^{ijl}x_j$, since the AH surface closely resembles a coordinate sphere. We present a more detailed discussion of the AHs in FLRW spacetimes  in appendix \ref{Sapp::AH}.

In particular, when focusing on the efficiency of the formation process, we will be interested in ratios  of the collapsing seed, i.e. $(\mbh/\Mxi)$, $(J_\mathrm{BH}/J^0_\xi)$ and $(a_\mathrm{BH} / \axi)$, where $\Mxi$ is the initial mass of the collapsing perturbation given by integrating the initial gradient and kinetic energy 
\begin{equation}
    \Mxi = \int dV~ \frac{1}{2}(\partial_i \xi)^2 +\Pi_\xi^2\,.\label{Seq::infall_mass}
\end{equation}
The dimensionful angular momentum $J_\xi^0$ is given by integrating the expression in \eqn{Seqn::angmomdensity_z} over volume on the initial slice and $\axi = J^0_\xi/\left(G\Mxi\right)^{2}$.

\section{Results} \label{Ssect:results}

\begin{figure*}[t]
    \includegraphics[width=\linewidth]{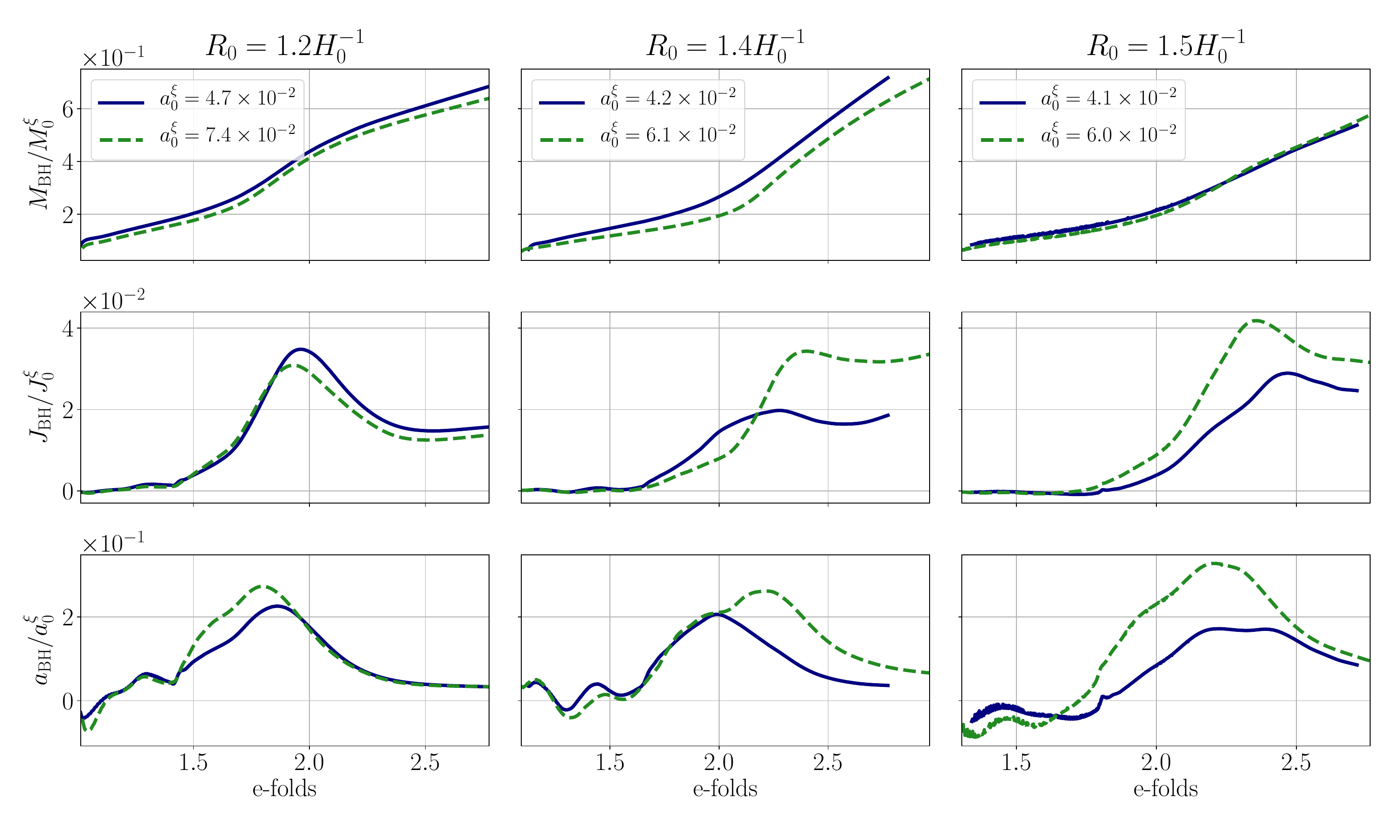}
    \vspace*{-10mm}
    \caption{\textbf{PBH masses, angular momenta and dimensionless spins for perturbations of varying initial size $R_0$.} The top, middle and bottom rows show normalized PBH mass $\mbh$, PBH dimensionful angular momentum $J_{\mathrm{BH}}$ and PBH dimensionless spin $a_{\mathrm{BH}}$ respectively, all as a function of number of e-folds after collapse initiated. The data in different columns is for shells of different initial size, as annotated at the top. The data is labelled by the shell's initial dimensionless spin. The legends in the top plots are valid for their entire respective columns. Our results show that the efficiency of the angular momentum transfer from shell to PBH is ${\cal O}(5)$\%.} 
    \label{Sfig:pbh_panel}
\end{figure*}

The length scale of interest in this work is the unperturbed Hubble horizon $H_0$, set through \eqn{Seq::initial_hubble_param}. For all simulations we choose $\phi_0 = 7.8\times 10^{-3} \mpl$ and the mass $m \approx 62 H_0$. We find that these values accurately model a matter-dominated universe expansion on average. For the initial size of the superhorizon perturbation we use a range of $R_0 \in [1.2, 1.5]H_0^{-1}$ with initial width  $\lambda = 0.15 H_0^{-1}$.  The spin amplitude and wavenumber are fixed to $B = 0.5$ and $k=2$, respectively.

In our previous work \cite{deJong:2021bbo} we found there exist two distinct mechanisms by which PBHs can form in a matter-dominated era depending on the amplitude of the initial perturbation, i.e.
\begin{equation}\label{Seq::direct_threshold}
    \frac{4 G \Mxi}{\lambda \left(2 + R_0 H_0\right)^2} \quad
    \begin{cases}
    < 1\quad \text{accretion collapse,}\\
    > 1\quad \text{direct collapse.}
    \end{cases}
\end{equation}

The two mechanisms are 
\begin{itemize}
    \item \textit{Accretion collapse}: The perturbation is not massive enough and disperses post-collapse, but triggers a gravitational seed that accretes the (non-rotating) background matter, which then collapses into a black hole -- {PBH formation dominated by $\phi$}.
    \item \textit{Direct collapse}: The initial perturbation is massive enough to directly collapse into a black hole -- {PBH formation dominated by $\xi$}. 
\end{itemize}
We find that this threshold accurately predicts whether direct or accretion collapse will take place, even when angular momentum is added. Given that the initial angular momentum is contained in the collapsing massless field $\xi$, we focus on the latter mechanism -- direct collapse. This is the most optimistic scenario to form PBHs with spin, as in the accretion case we expect most of the angular momentum of $\xi$ to be radiated away when the initial perturbation disperses\footnote{The angular momentum transfer between the fields $\xi$ and $\phi$ is minimal if they only couple via gravity.}.
Consequently, for the perturbation's radial amplitude we use a range of $A \in [0.0825, 0.09]\mpl$. 
We vary $\omega \in [0, 12] H_0^{-1}$ to parameterize the amount of angular momentum in the system. 

\subsection{PBH mass} \label{Ssect::results_pbhmass}

We show PBH mass data for perturbations of three different initial sizes in the top row of Fig. \ref{Sfig:pbh_panel}. We conclude that the efficiency is $\sim 15\%$ and is only weakly sensitive to the initial angular momentum and size of the perturbation. Naively, one might expect that higher initial angular momentum will prevent matter from collapsing, as inward gravitational acceleration can to an extent be balanced by rotational motion, so an increased angular momentum could result in a decrease in the efficiency $(\mbh/\Mxi)$. Our results show that this effect is not dominant -- we suspect that this is related to the fact that much of the perturbation's angular momentum is concentrated away from the equatorial plane (unlike, say, that of an accretion disk), and this spinning matter does not orbit but instead spirals into the centre. We comment further on this effect in appendix \ref{Sapp::initial_data}.

Furthermore, we note that the mass accretion  \emph{rate} of the PBHs is equally insensitive to the angular momentum of the perturbation, at least for the spin values we probe. This means that the accreted mass can quickly surpass the initial seed mass, so that predictions for final PBH mass will depend heavily on assumptions made about the continued accretion rate. We note that the dependence of the accretion rate on the perturbation's angular momentum may well change if the initial angular momentum increases considerably, which is an interesting direction for future studies.

\subsection{PBH spin}\label{Ssect::results_pbhspin}

We investigate the evolution of the dimensionful PBH angular momentum around the $z$-axis $J_{\textrm{BH}}$. We plot the corresponding efficiency ($J_\mathrm{BH}/J^0_\xi$) in the middle row of Fig. \ref{Sfig:pbh_panel}, from which we see that $J_\mathrm{BH}$ is consistent with zero initially, meaning the non-spinning parts of the perturbation cause initial AH formation. The PBH then starts accreting matter with angular momentum until $J_\mathrm{BH}$ peaks at $2-4\%$ of the initial total angular momentum $J^0$ and finally asymptotes to a constant value.

There is a post-peak dip in angular momentum that appears consistently for all simulations, which we believe are the result of some components of the shell spinning up, meaning local angular momentum conservation requires other parts to spin in the opposite direction. This causes the formation of regions with net negative spin, whose accretion by the PBH causes the dip.

Finally, we show the efficiency of the dimensionless spin ($a_\mathrm{BH}/\axi$) in the bottom row of Fig. \ref{Sfig:pbh_panel}, which peaks at a maximum $\sim 25\%$. Overall, our results suggest that the peak efficiencies $(J_\mathrm{BH}/J^0_\xi)$ and $(a_\mathrm{BH} / \axi)$ increase with increasing perturbation radius, and whether or not this trend continues for even larger radii and values of $\axi$ needs to be investigated further. Lastly, $a_\mathrm{BH}$ decreases very quickly due to the accretion of the background matter, which is non-rotating. This demonstrates post-formation evolution and dynamics are important.
\section{Discussion} \label{Ssect:conclusions}

In this work, we use numerical relativity to show that sufficiently massive superhorizon perturbations with inherent angular momentum will generically collapse into a PBH with spin. This process is rather efficient: $\sim 10\%$ of the initial mass and $\sim 5\%$ of the initial angular momentum make up the PBH at AH formation. We show that the initial PBH mass and immediate post-formation accretion rate only depend weakly on the perturbation's initial size and angular momentum, for the parameters we explore. 
The PBH  spin $(a_\mathrm{BH}/\axi)$ efficiency peaks at $\sim 25\%$ but crucially decreases quickly during the subsequent evolution as during a matter-like era, black holes keep growing as they accrete non-rotating background matter. 

To illustrate this, we assume that the rapid accretion rate shown in Fig. \ref{Sfig:pbh_panel} levels off quickly and that the PBH continues to grow self-similarly\footnote{We emphasize that instead, the rapid accretion rate may be sustained until the PBH mass is proportional to the Hubble mass. Additionally, subsequent growth could be significantly slower than self-similar.}, i.e. $\mbh\propto 1/H$, so that the dimensionless spin evolves as $a_\mathrm{BH}\propto H^2$. Using the matter-dominated era scaling between the Hubble parameter and the temperature $H\propto T^{3/4}$, we obtain
\begin{equation}
    \frac{a_\mathrm{BH}(T)}{a_\mathrm{BH}(T_0)} \approx \left(\frac{T}{T_0}\right)^{3/2},
\end{equation}
where $a_\mathrm{BH}(T)$ is the dimensionless black hole spin at temperature $T$, and $T_0$ is the temperature at formation. In the most optimistic scenario where the PBH is the endpoint of a highly rotating initial seed and is near extremal $a_\mathrm{BH}^0\approx 1$, this implies that the typical leftover spin is $\leq \mathcal{O}(0.1)$ if the duration of the matter-dominated epoch is $\Delta T \gtrapprox 0.8T_0$. 

Computational cost limits us from tracking the PBH evolution beyond a few e-folds after formation, so the assumptions we make about the continued accretion rate are important. However, PBHs are expected to accrete significantly during a matter-dominated epoch, even if the exact rate at which they do so is unknown. Therefore, we argue that if PBHs with large spins are to form in a matter-dominated epoch, one needs two ingredients: firstly, the PBHs must be highly spinning at formation and secondly, the matter-dominated epoch's duration must be short.

\acknowledgments

We would like to thank the GRChombo Collaboration team \href{http://www.grchombo.org}{(http://www.grchombo.org/)} and the COSMOS team at DAMTP, Cambridge University for their ongoing technical support. JCA acknowledges funding from the Beecroft Trust and The Queen’s College via an extraordinary Junior Research Fellowship (eJRF). 

This work was performed using the DiRAC@Durham facility managed by the Institute for Computational Cosmology on behalf of the STFC DiRAC HPC Facility (www.dirac.ac.uk) under DiRAC RAC13 Grant ACTP238 and DiRAC RAC15 Grant ACTP316. The equipment was funded by BEIS capital funding via STFC capital grants ST/P002293/1, ST/R002371/1 and ST/S002502/1, Durham University and STFC operations grant ST/R000832/1. This work also used the DiRAC Data Intensive service at Leicester, operated by the University of Leicester IT Services, which forms part of the STFC DiRAC HPC Facility (www.dirac.ac.uk). The equipment was funded by BEIS capital funding via STFC capital grants ST/K000373/1 and ST/R002363/1 and STFC DiRAC Operations grant ST/R001014/1. DiRAC is part of the National e-Infrastructure.

\bibliography{mybib}

\newpage
~\newpage
\appendix

\counterwithin{figure}{section}

\section{Apparent horizons in FLRW spacetimes}\label{Sapp::AH}

To find the AH of the formed PBH, we follow the procedure described in \cite{Thornburg:2003sf}. The AH is the outermost marginally outer-trapped surface on which the expansion of outgoing null geodesics $\Theta_+$ vanishes, i.e.
\begin{equation} \label{Spos_expansion_zero}
    \Theta_+\equiv D_i s^i + K_{ij} s^i s^j - K = 0~.
\end{equation}

In FLRW spacetimes one often encounters cosmological horizons, for which a local measure can be found in a similar manner using the expansion of ingoing null vectors $k_-^a = n^a - s^a$, equivalent to
\begin{equation} \label{Sneg_expansion_zero}
    \Theta_-\equiv -D_i s^i + K_{ij} s^i s^j - K = 0~.
\end{equation}
In the following, we will refer to $\Theta_+$ ($\Theta_{-}$) as outgoing (ingoing) expansion, and we note that it measures the fractional change in the area of an outward (inward) spherical flash of light \cite{baumgarte_shapiro}.

In general, these equations must be solved numerically, using a nonlinear root finder algorithm such as the Newton-Raphson method or quasi-Newton methods. However, to build some intuition for these quantities, it is useful to consider specific cases with a high degree of symmetry, which can be solved analytically. It can be shown \cite{baumgarte_shapiro} that for a spherically symmetric line element of the form
\begin{equation}
    ds^2 = -\alpha^2 dt^2 + \psi(t,r)^2 \left[dr^2 +r^2\left( d\theta^2+\sin^2\theta\, d\phi^2\right)\right]~,\nonumber
\end{equation}
and with $s^r=1/\psi(t,r)$, $s^\theta=s^\phi=0$, the outgoing/ingoing expansion simplifies to
\begin{equation}
    \Theta_\pm = \frac{2}{\psi}\left[\frac{\partial_t\psi}{\alpha} \pm \left(\frac{\partial_r\psi}{\psi} + \frac{1}{r}\right)\right].
\end{equation}
We find it instructive to treat some concrete examples here and we discuss the resulting expressions for a Schwarzschild, FLRW and McVittie spacetime below. We will only consider positive radius solutions to \eqn{Spos_expansion_zero} and \eqn{Sneg_expansion_zero}, regarding negative radius solutions as unphysical.

\begin{itemize}
    \item[(i)] For a Schwarzschild black hole, the conformal factor in isotropic coordinates is $\psi_\mathrm{BH}(r) = \left(1+GM/2r\right)^2$ and the lapse is $\alpha_\mathrm{BH}(r)=(1-GM/2r)/(1+GM/2r)$. The expansions are then given by
    \begin{equation}
        \Theta_\pm=\mp\frac{8\left(GM-2r\right)r}{(GM+2r)^3}~,
    \end{equation}
    which both vanish at $r=0$ and at the black hole horizon $r=GM/2$.
    
    In these coordinates, the areal radius for the Schwarzschild metric is given by $r \psi_\mathrm{BH}(t, r)$ and therefore, the area of a spherical surface is $A = 4\pi (r \psi_\mathrm{BH}(t, r))^2$ \footnote{We note that at $r=GM/2$ and with units restored, this area equals $16\pi G^2M^2$, as it should for this spacetime.}. The outgoing (ingoing) expansion can be thought of as measuring the fractional change in the area of an outward (inward) spherical flash of light. From the expression for $A$, this area decreases (increases) inside the apparent horizon and increases (decreases) outside the apparent horizon when the coordinate radius is increased (decreased). This justifies the fact that $r=GM/2$ solves both $\Theta_\pm = 0$.
    
    \item[(ii)] In an FLRW spacetime sliced by cosmic time $t$, $\psi_\mathrm{FLRW}(t) = a(t)$ and $\alpha_\mathrm{FLRW}=1$, so that 
    \begin{equation}
        \Theta_\pm=2\left(H \pm \frac{1}{ar}\right)~,
    \end{equation}
    and $\Theta_-$ vanishes at the comoving Hubble horizon $r=a^{-1}H^{-1}$. This is the point at which the universe begins to expand superluminally relative to the origin, and ingoing rays, that converge to the origin for smaller radii, end up receding from us.
    
    \item[(iii)] A black hole immersed in an FLRW background can be represented by the McVittie line element \cite{McVittie:1933zz}, for which $\alpha = \alpha_\mathrm{BH}\left(ar\right)$ and $\psi = \psi_\mathrm{FLRW}(t)\psi_\mathrm{BH}\left(ar\right)$, so that
    \begin{equation}
        \Theta_\pm=2\left(H\mp\frac{4ar\left(GM-2ar\right)}{\left(GM+2ar\right)^3}\right)~.
    \end{equation}
    When $GMH\ll 1$, as is the case in the situations we consider, $\Theta_-$ vanishes for $$r\approx\frac{1}{aH} - \frac{2GM}{a} \quad \text{and} \quad r\approx\frac{GM}{2a} + \frac{2H (GM)^2}{a}\,,$$ while $\Theta_+$ vanishes for $$r\approx\frac{GM}{2a} - \frac{2H (GM)^2}{a} \quad \text{and} \quad r\approx \frac{H(GM)^2}{4a}\,.$$ This means that an AH finder should in principle be able to find the PBH and cosmological horizon, as well as a perturbed $r=0$ solution. This is confirmed by our simulations. We are able to distinguish the two solutions for $\Theta_+ = 0$ because the physical area of the latter shrinks, rendering it unphysical.
\end{itemize}

\section{Initial data} \label{Sapp::initial_data}
\begin{figure*}[t]
    \includegraphics[width=\linewidth]{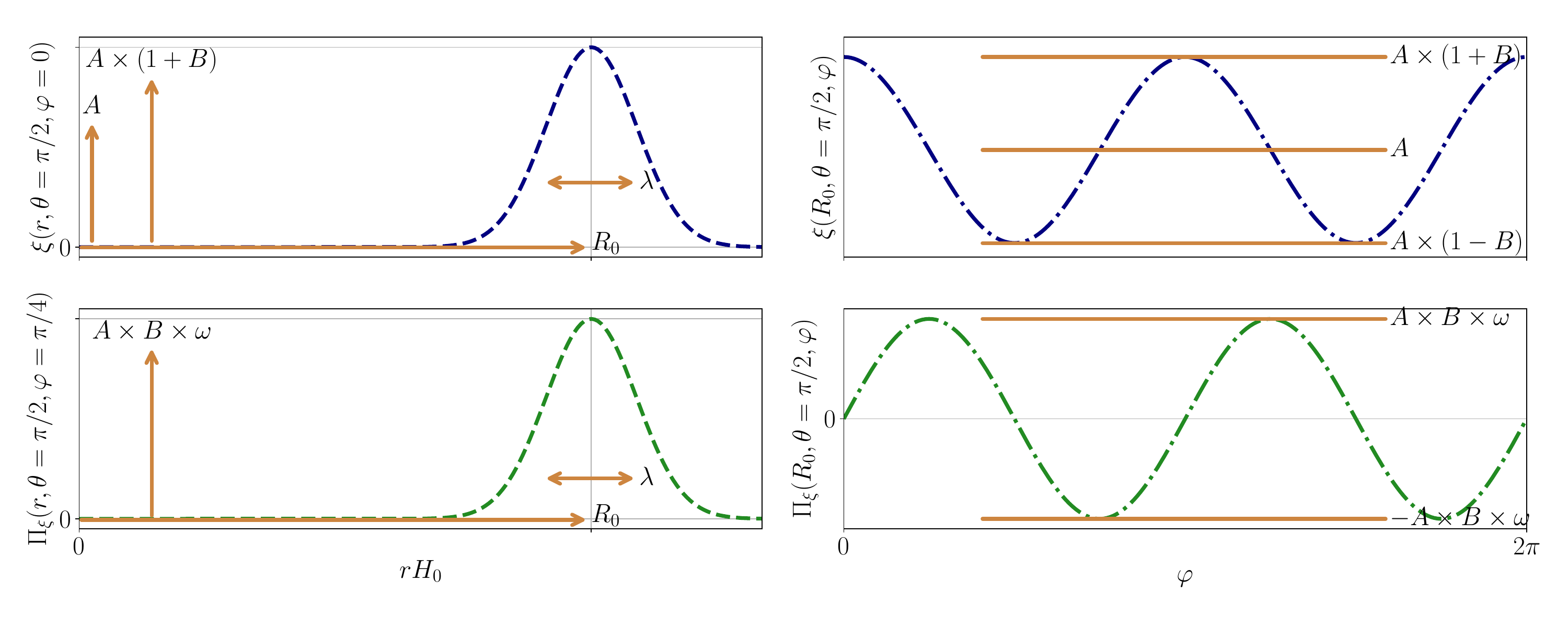}
    \vspace*{-10mm}
    \caption{\textbf{Initial configurations for $\xi$ and $\Pi_\xi$} give by \eqn{Seq::xi_init_time} and \eqn{Seq::pi_init_time}, for parameters $A=0.0825\mpl, R_0 = 1.2H_0^{-1}, \lambda = 0.15H_0^{-1}, B = 0.5, k = 2, \omega = 12H_0^{-1}$. The navy blue (green) dashed lines in the top (bottom) row represent initial profiles for $\xi$ ($\Pi_\xi$). The dashed (dashdotted) lines in the left (right) columns represent initial $\xi$ or $\Pi_\xi$ profiles as a function of coordinate radius $r$ with elevation angle $\theta$ and azimuthal angle $\varphi$ kept constant (as a function of $\varphi$ with $r$ and $\theta$ kept constant). The various solid orange arrow and lines show how the constants $A, R_0, \lambda, B, \omega$ relate to the initial shape of these profiles, whilst the final constant $k$ is set to $2$ here, so that the profiles in the right column oscillate twice per $\varphi$ rotation.} 
    \label{Sfig:init_params}
\end{figure*}

\begin{figure}[t]
    \includegraphics[width=\linewidth]{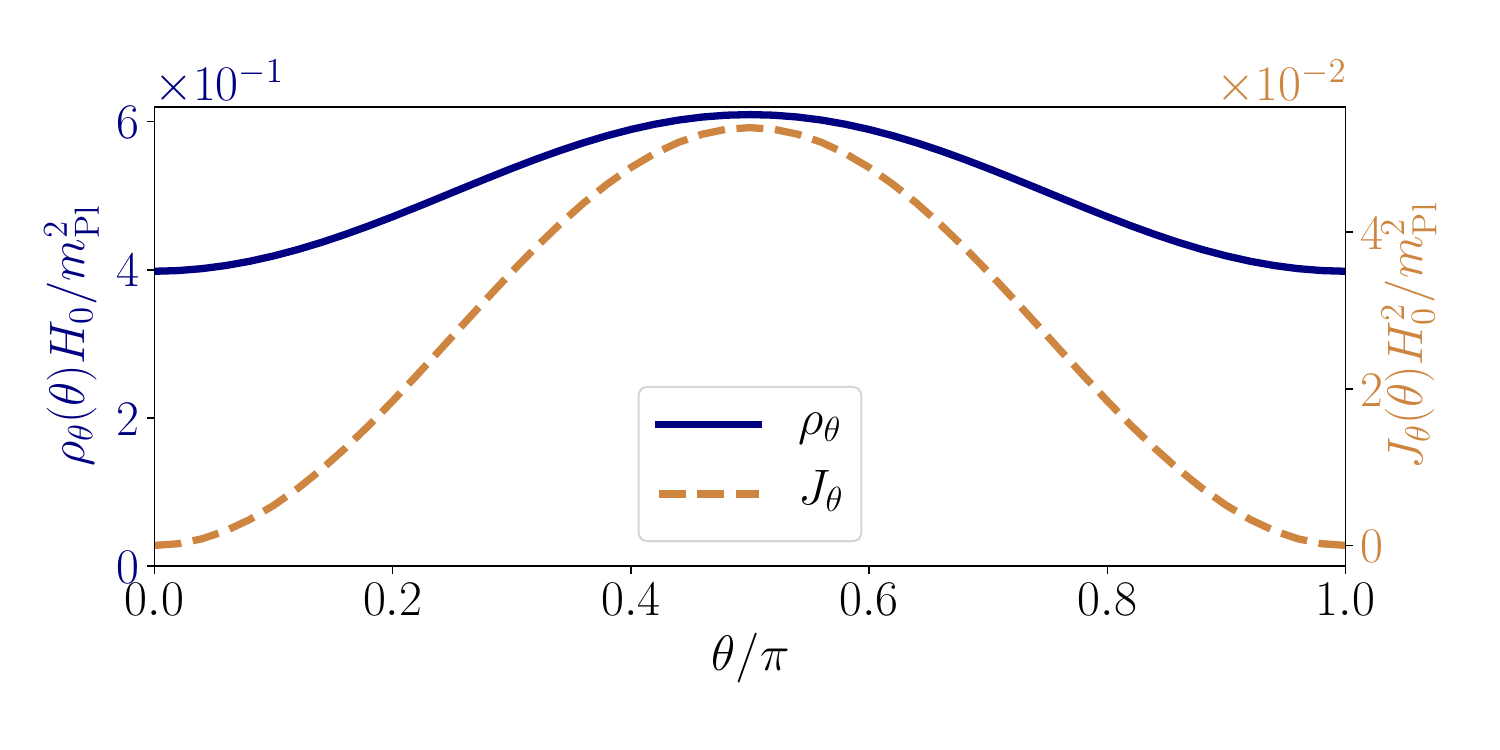}
    \vspace*{-10mm}
    \caption{\textbf{Initial configurations of $\rho_\theta(\theta)$ and $J_\theta(\theta)$,} defined in \eqn{Seqn::rhotheta} and \eqn{Seqn::Jtheta}, as a function of elevation angle $\theta$, for initial parameters $A = 0.0875\mpl$, $R_0 = 1.4H_0^{-1}$, $\lambda = 0.15H_0^{-1}$, $B = 0.5$, $k = 2$, $\omega = 12H_0^{-1}$. The solid blue (dashed light brown) line corresponds to the y-axis on the left (right).} 
    \label{Sfig:theta_plot}
\end{figure}

The initial configurations for $\xi$ and its time derivative $\Pi_\xi$ are described in section \ref{Ssect::init_data} of the main text and illustrated in Fig. \ref{Sfig:init_params}. 

In section \ref{Ssect::results_pbhmass} of the main text, we show that the PBHs' initial masses and accretion rates do not decrease significantly with increasing initial shell angular momentum and we attribute this to the initial shape of the shell. In Fig. \ref{Sfig:theta_plot}, we show typical initial profiles of the quantities

\begin{subequations}
\begin{align}
    \rho_\theta(\theta) &= \iint ~dr d\alpha \hspace{1mm} r^2 \rho(r, \theta, \alpha) \label{Seqn::rhotheta}, \\
    J_\theta(\theta) &= \iint ~drd\alpha \hspace{1mm} r^2 \mathcal{J}(r, \theta, \alpha), \label{Seqn::Jtheta}
\end{align}
\end{subequations}
i.e. the energy density and angular momentum density integrated over a conical surface at elevation angle $\theta$. From this figure, we note that the largest part of the shell's mass is located away from the equatorial $z=0$ plane, where it is less susceptible to orbital effects and more likely to spiral into the centre. Additionally, a significant part of the angular momentum is located away from the equatorial plane. We expect that if the mass distribution peaked more sharply around the equatorial plane, the spin effects on the initial PBH mass and accretion rate would be stronger. This is an interesting direction for future research. 

\section{Convergence testing} \label{Sapp::convergence_testing}
\begin{figure*}[t]
    \includegraphics[width=\linewidth]{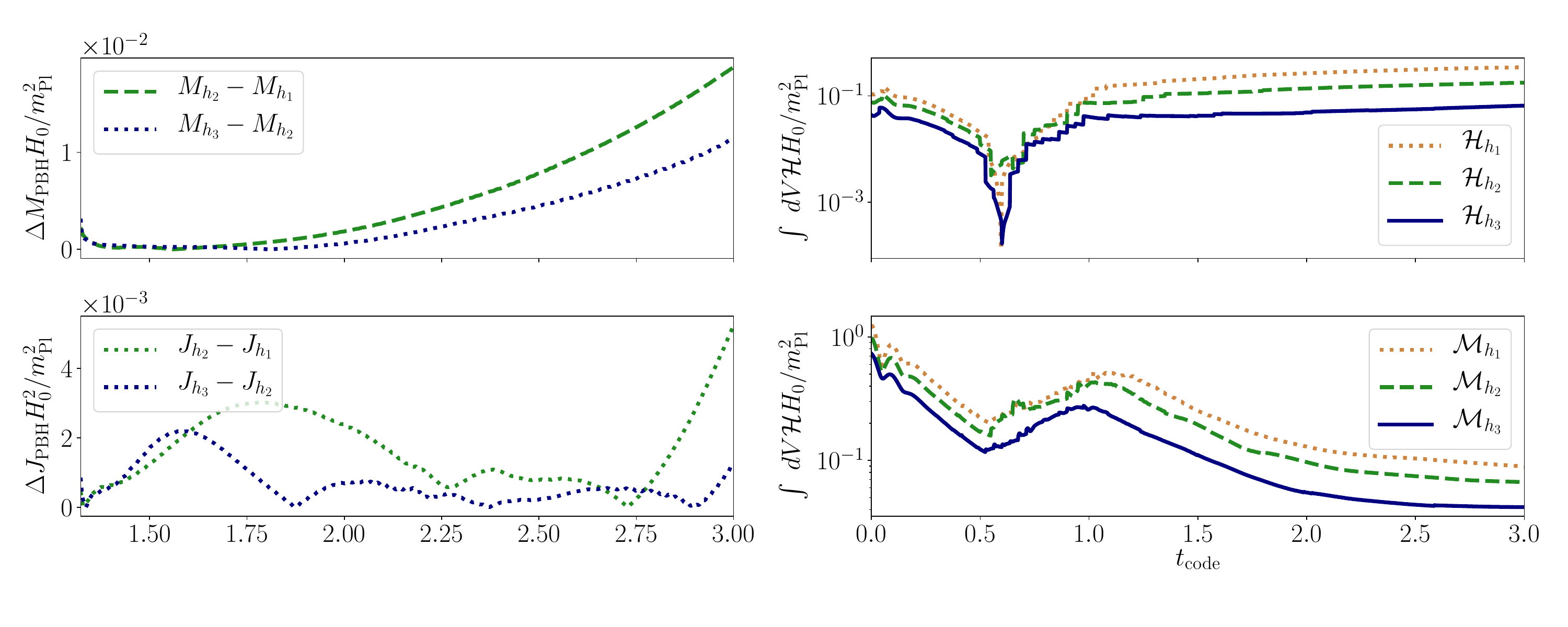}
    \vspace*{-10mm}
    \caption{\textbf{Convergence testing.} We show data for the formation of a PBH formed by an initial perturbation with $R_0 = 1.4$ and $\chi_0^\xi = 6.1\times 10^{-2}$. The top left panel shows the absolute value of the difference between the masses detected by simulations of different resolutions, while the bottom left panel shows the absolute value of the difference between the dimensionless spins. The top right panel shows the Hamiltonian constraint violation integrated over the simulation domain for different resolutions, while the bottom right panel shows the integrated momentum constraint violation.} 
    \label{Sfig:convergence}
\end{figure*}

We perform several convergence tests on the robustness of our numerical results, by finding the mass and spin of a PBH formed by an initial perturbation with $R_0 = 1.4H_0^{-1}$ and $a_\xi^0 = 6.1\times 10^{-2}$. We do so using three different base grid resolutions, $N_1 = 80$, $N_2 = 96$ and $N_3 = 128$, which correspond to base grid spacings $h$ of $h_1 = 6.25\times 10^{-2}H_0^{-1}$, $h_2 = 5.21\times 10^{-2} H_0^{-1}$ and $h_3 = 3.90\times 10^{-2} H_0^{-1}$. We track Hamiltonian and momentum constraint violation, as well. 

Fig. \ref{Sfig:convergence} shows mass and spin difference between simulations of different base resolutions and Hamiltonian and momentum constraint violation for all simulations, indicating that convergence is achieved. We note that we checked the code simulates a homogeneous FLRW universe for appropriate initial conditions in an earlier publication \cite{deJong:2021bbo}.

\section{Evolution equations}\label{Sapp::evolution_equations}

Whilst the metric variable evolution equations can be found in e.g. \cite{Alic_2012}, we list the matter evolution equations here explicitly for completeness. For $N$ minimally coupled scalar fields $\phi_i$, the evolution equations are 
\begin{subequations}
\begin{align}
    \partial_t \phi_i &= \alpha \Pi_i + \beta^j\partial_j \phi_i, \\
    \partial_t \Pi_i &= \beta^j\partial_j \Pi_i + \alpha\partial^j\partial_j \phi_i + \partial^j \phi_i \partial_j \alpha\\ 
    &\quad+ \alpha\left(K\Pi_i - \gamma^{jk}\Gamma^l_{jk}\partial_l\phi_i - \frac{dV(\phi_i)}{d\phi_i}\right)\nonumber,
\end{align}
\end{subequations}
whilst the energy-momentum expressions appearing in the metric variables' evolution equations are
\begin{subequations}
\begin{align}
    \rho &= \frac{1}{2}\sum_{i=1}^N \left[\Pi_i^2 + \partial^j\phi_i\partial_j\phi_i + V(\phi_i)\right],\\
    S_j &= \sum_{i=1}^N -\Pi_i \partial_j\phi_i,\\
    S_{jk} &= \sum_{i=1}^N\Big[\partial_j\phi_i\partial_k\phi_i - \frac{1}{2}\bar{\gamma}_{jk}\big(\bar{\gamma}^{lm}\partial_l\phi_i\partial_m\phi_i\\
    &\quad - \Pi_i^2 - 2V(\phi_i) \big) \Big],\nonumber\\
    S &= \gamma^{jk}S_{jk}.
\end{align}
\end{subequations}

\end{document}